\documentclass[prb,amsmath,amssymb,floatfix,twocolumn,superscriptaddress]{revtex4}
\usepackage{bm}
\usepackage{url}
\usepackage{graphicx}
\usepackage{physics}
\usepackage{epsfig}
\usepackage{subfigure}
\usepackage[usenames]{color}
\usepackage{hyperref}
\usepackage[normalem]{ulem}

\begin{document}

\title{Ab Initio Mismatched Interface Theory of Graphene on $\alpha$-RuCl$_3$: Doping and Magnetism}

\author{Eli Gerber}
\affiliation{School of Applied and Engineering Physics, Cornell University, Ithaca, New York 14853, USA}
\author{Yuan Yao}
\affiliation{Department of Physics, Cornell University, Ithaca, New York 14853, USA}
\author{Tomas A. Arias}
\affiliation{Department of Physics, Cornell University, Ithaca, New York 14853, USA}
\author{Eun-Ah Kim}
\affiliation{Department of Physics, Cornell University, Ithaca, New York 14853, USA}

\begin{abstract}
Recent developments in twisted and lattice-mismatched bilayers have revealed a rich phase space of van der Waals systems and generated excitement. Among these systems are heterobilayers which can offer new opportunities to control van der Waals systems with strong in plane correlations such as spin-orbit-assisted Mott insulator $\alpha$-RuCl$_3$.
Nevertheless, a theoretical {\em ab initio} framework for mismatched heterobilayers without even approximate periodicity is sorely lacking.  We propose a general strategy for calculating electronic properties of such systems, 
mismatched interface theory (MINT), and apply it to the graphene/$\alpha$-RuCl$_{3}$ (GR/$\alpha$-RuCl$_{3}$) heterostructure.  
Using MINT, we predict uniform doping of 4.77\% from graphene to $\alpha$-RuCl$_3$ and magnetic interactions in $\alpha$-RuCl$_3$ to shift the system toward the Kitaev point. Hence we demonstrate that MINT can guide
 targeted materialization of desired model systems and 
discuss recent experiments on GR/$\alpha$-RuCl$_{3}$ heterostructures.

\end{abstract}

\date{\today \ [file: \jobname]}

 \maketitle

New capabilities for synthesizing atomic scale heterostructures with lattice-mismatched van der Waals materials have opened the floodgates to an infinite array of possibilities.
Among them are twisted structures of identical monolayers, such as multi-layer graphene\cite{Pablo2018,Young-Dean} and transition metal dichalcogenides\cite{PhysRevB.93.180501} (TMDs), as well as structures involving two distinct monolayers, such as TMD heterobilayers\cite{MoS2WSe2,MoS2WS2} and the GR/$\alpha$-RuCl$_{3}$ heterostructure\cite{Henriksen,Mashhadi}. These capabilities offer a new control parameter to design new systems. Unfortunately, traditional \emph{ab initio} techniques for calculating the electronic structure of materials are powerless when the lattice mismatch between two crystals leads to the absence of periodicity \cite{incom1,incom2}. 

For twisted graphene bilayers, moir\'{e} materials offer a superlattice and the community poured on theoretical efforts to construct effective tight-binding models \cite{band-MacDonald,band-Kaxiras,band-Sentil-PhysRevB.98.085435,Fu-PhysRevB.98.045103,massatt2017electronic,massatt2018incommensurate} and develop specialized techniques for solving those models in extended, aperiodic systems\cite{band-Kaxiras,carr2017twistronics}. Pioneering efforts have also been made to develop perturbation theory for interlayer coupling that is homogenized in the in plane directions\cite{incom2}, though actual calculations within this latter approach also have required the use of non-self-consistent tight-binding models. Without self-consistency, such tight-binding-based approaches are limited to homobilayers in that they cannot account for effects such as screening and charge transfer.

\begin{figure}[t]
\begin{centering}
\includegraphics[width=.5\textwidth]{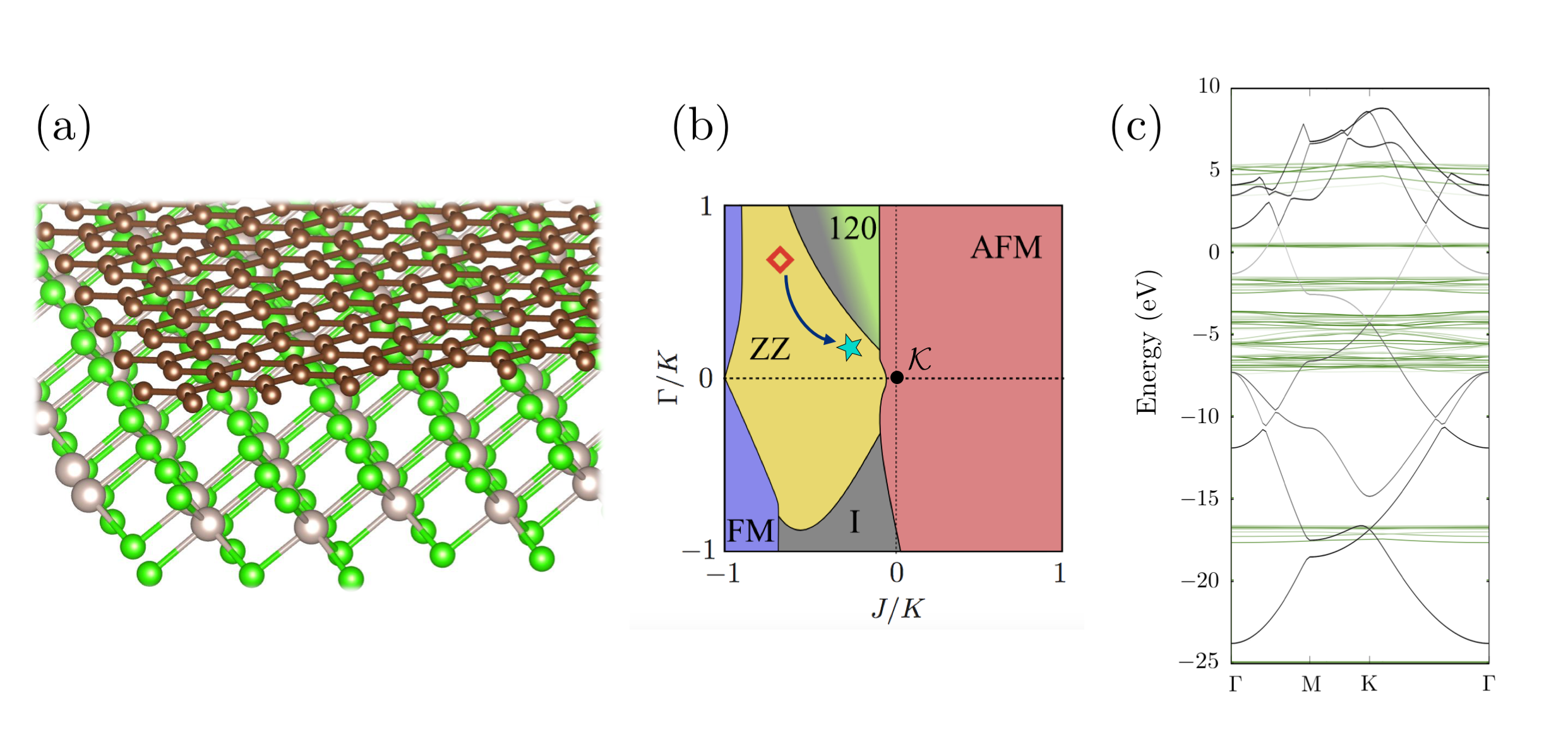}
\end{centering}
\caption{(a) A GR/$\alpha$-RuCl$_{3}$ bilayer. 
(b) The MINT results for our GR/$\alpha$-RuCl$_3$ system (green star) added to the Luttinger-Tisza phase diagram from Ref.~\cite{Kim-Kee-PD-PhysRevB.91.241110} with the red diamond representing the ground state of plain  RuCl$_{3}$. ``ZZ" denotes the zigzag antiferromagnetic phase, ``AFM" is the regular antiferromagnet, ``FM" is the ferromagnet, ``120" is the 120$^{\circ}$-ordered phase, and ``I" is the incommensurate order. The Kitaev point at the origin is denoted $\mathcal{K}$. (c) Calculated band structures of graphene (black and gray) and $\alpha$-RuCl$_3$ (green).}
\label{fig:ideal}
\end{figure}

On the other hand, heterobilayers without superlattice structure are lacking a theoretical framework outside of making supercells\cite{LAZIC2015324}, which are computationally costly and introduce significant strain. Of our particular interest is the  GR/$\alpha$-RuCl$_{3}$ system. $\alpha$-RuCl$_{3}$ (RuCl$_{3}$) is a layered, spin-orbit-assisted Mott insulator that lies very close to forming the exotic quantum spin liquid ground state\cite{YJKim2014,Kim-scattering-PhysRevLett.114.147201,Field-PhysRevLett.119.037201,Ji-NP2017,Zheng-NMR-PhysRevLett.119.227208,inelastic-Banerjee2017,field-NMR-Ruegg-2018,kappa-Matsuda2018}. Hence heterostructuring may offer a tantalizing possibility of exploring the phase diagram and doping the quantum spin liquid. However the large mismatch (see Fig.\ref{fig:ideal}(a)) rules out meaningful superlattice formation and the work-function difference ($\phi_{r}=6.1$ eV for $\alpha$-RuCl$_3$\cite{rucl3work} and $\phi_{g}=4.6$ eV for graphene\cite{graphenework}) suggests charge transfer. 

We introduce a new framework for fully self-consistent electronic-structure studies of
lattice-mismatched atomic heterostructures called mismatched interface theory (MINT). We then apply this approach to carry out full, direct density-functional theory studies of the GR/RuCl$_{3}$ heterostructure and predict that electrons from the graphene layer dope RuCl$_{3}$ while moving the system closer to the Kitaev point\cite{KITAEV20062} [see Fig.~\ref{fig:ideal}(b)]. RuCl$_{3}$ has been hypothesized to lie in the spin liquid regime at this phase point\cite{Kim-Kee-PD-PhysRevB.91.241110}, where spin correlation functions beyond nearest-neighbor (NN) vanish and, on a given NN bond, only the components of spins matching the bond type remain correlated\cite{PhysRevLett.98.247201}.

Pure RuCl$_{3}$ has been intensely studied since it was recognized to be a candidate system to materialize the honeycomb lattice Kitaev model\cite{KITAEV20062} with extremely anisotropic spin-spin interaction among $j_{eff}=1/2$ pseudospin moments on the Ruthenium sites. In the bulk crystal the 
edge-sharing
RuCl$_6$ octahedra form two-dimensional RuCl$_3$ layers with weak interlayer van der Waals coupling. Although there are signs of Kitaev physics in the bulk system, it orders into a zigzag antiferromagnet at $T_{Neel}$\cite{Kim-scattering-PhysRevLett.114.147201,Field-PhysRevLett.119.037201,Ji-NP2017,Zheng-NMR-PhysRevLett.119.227208,inelastic-Banerjee2017,field-NMR-Ruegg-2018} and evidence of Kitaev quantum spin liquid physics is only seen at temperatures above the ordering temperature\cite{Ji-NP2017,inelastic-Banerjee2017} or under a magnetic field, which suppresses ordering\cite{Field-PhysRevLett.119.037201,field-NMR-Ruegg-2018,Zheng-NMR-PhysRevLett.119.227208,field-NMR-Ruegg-2018,kappa-Matsuda2018}. 
\textcite{Kim-Kee-PD-PhysRevB.91.241110} obtained an effective model that captures competing interactions leading to the zigzag and other nearby orders from {\it ab initio} studies: \footnote{Here we are keeping only the nearest neighbor coupling as further neighbor coupling constants were found to be suppressed by over 200\%.} 
\begin{equation}
H_{J K \Gamma}=\sum_{\langle ij \rangle \in \alpha \beta (\gamma)} \big[  K S_{i}^{\gamma}S_{j}^{\gamma} + J \mathbf{S}_{i} \cdot \mathbf{S}_{j} +\Gamma(S_{i}^{\alpha}S_{j}^{\beta} + S_{i}^{\beta}S_{j}^{\alpha}) \big],
\label{eq:RuCl3-H}
\end{equation}
where $i,$ $j$ designate the Ru$^{3+}$ sites and the $S_{i}^{\alpha}$ are components of the $j_{eff}=1/2$ pseudospin operator $\mathbf{S}_{i}$, $\alpha \beta (\gamma)$ labels a bond on which the spin direction $\gamma$ is fixed. 
They further placed the model parameters relevant for RuCl$_3$ in the zigzag-ordered phase close to the ferromagnetic- and $120^\circ$-ordered state in a classical phase diagram based on Luttinger-Tisza analysis\cite{Luttinger-Tisza-PhysRev.70.954} [see the reproduced phase diagram in Fig.~\ref{fig:ideal}(b)]. 
However, little is known about how to move the system closer to the Kitaev point at the origin. Here we will use MINT to extract the tight-binding parameters for the graphene-RuCl$_3$ heterostructure to obtain the Kitaev ($K$), Heisenberg ($J$) and symmetric off-diagonal exchange coupling ($\Gamma$) constants.

\emph{Mismatched interface theory (MINT) ---} Electronic structure theory offers two broad sets of approaches for treatment of either isolated or periodic systems, respectively. Standard praxis for treating periodic structures within isolated-system methods is to construct large clusters of periodic material. Conversely, to treat aperiodic structures within periodic methods, one constructs large, periodic ``supercells'' containing the aperiodic structure. Both methods depend on the nearsightedness of electronic matter (NEM) \cite{kohn1996density,prodan2005nearsightedness} to ensure convergence toward exact behavior
as the size of the calculation increases to infinity. This well-established principle is reflected in
the mathematics underlying 
the recently developed tight-binding-based methods for twisted bilayers\cite{massatt2017electronic,carr2017twistronics}. 

\begin{figure}
    \centering
    \includegraphics[height=.2\textwidth]{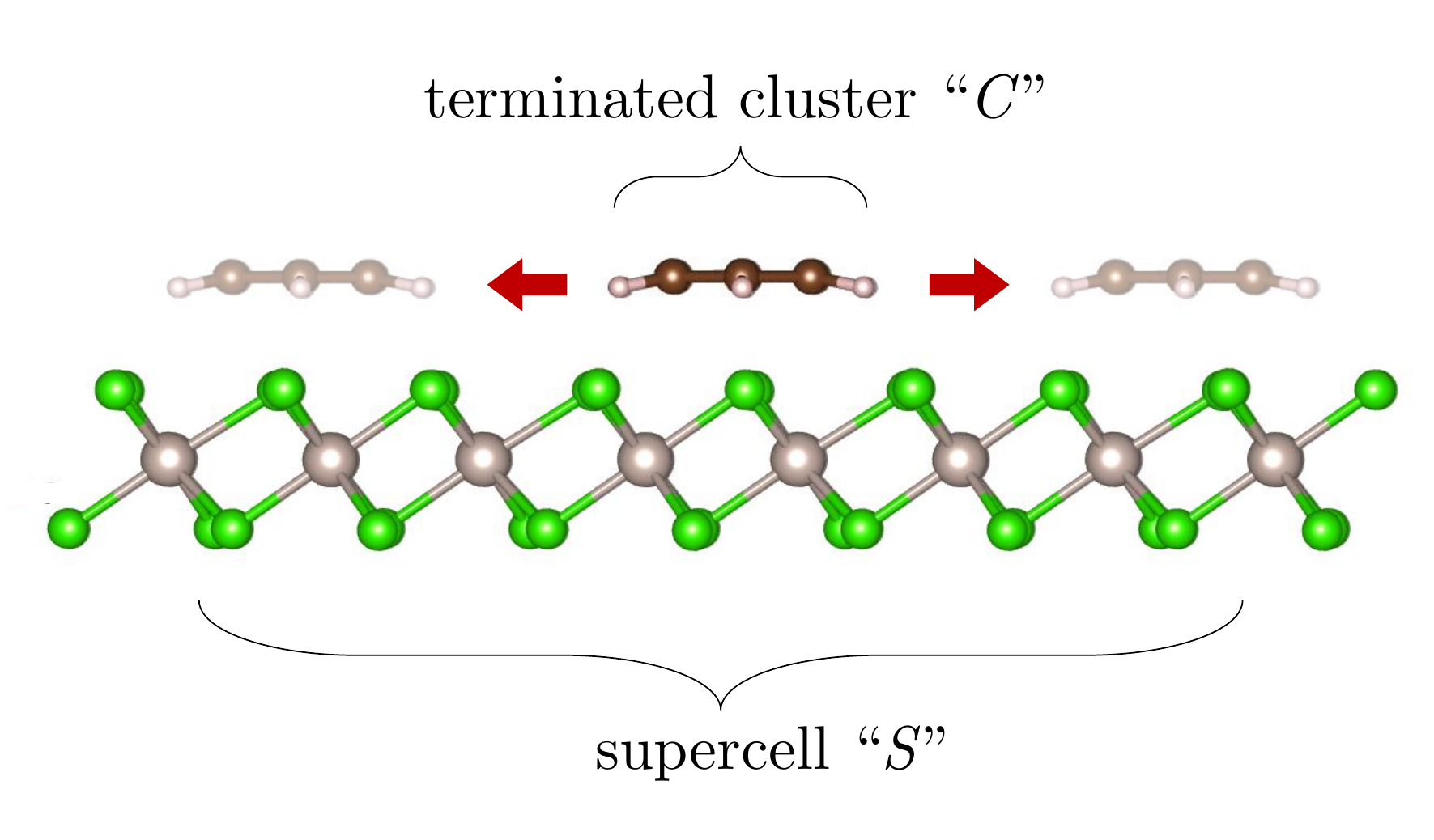}
    \caption{The terminated cluster $C$ and the supercell $S$. }
    \label{fig:MINT}
\end{figure}

We here demonstrate for the first time that a simple combination of the supercell and cluster approaches allows treatment of incommensurate interfaces directly with standard density-functional theory software \textit{without the need for specialized techniques} or reduction to non-self-consistent tight-binding models. We find, moreover, that nearsightedness ensures sufficiently rapid convergence that the calculations for our system of interest are quite practical.
The basic approach, illustrated in Fig.~\ref{fig:MINT}, begins with a large, periodic supercell of the material system of primary interest $S$ (\textit{e.g.}, single-layer $\alpha$-RuCl$_3$.) Next, we place into this supercell clusters $C$ of the material from the second subsystem, with terminating groups such that no bonds are left free and adjacent cluster images do not interact (\textit{e.g.}, hydrogen-terminated graphene flakes).

Finally, we study convergence as the cluster size is increased.  Nearsightedness then ensures that sufficiently far from the boundary of the cluster, both materials behave just as they would for a truly infinite interface. Moreover, as the cluster grows, it eventually samples all possible registries with the other material.  Appropriate finite-size scaling of thermodynamically intrinsic quantities then enables extraction of the behavior of the infinite interface. As a matter of practice, following past tight-binding work\cite{carr2017twistronics}, convergence with respect to sampling over registries (and even local rotational disorder) can be accelerated in systems with large moir\'{e} patterns by employing smaller clusters and averaging over different relative positions and orientations. Below, we present the first fully self-consistent density-functional theory (DFT) calculations carried out within this approach. 

The final step of MINT is to use the ``MINT representation'' to make predictions for the electronic structure and magnetic interactions of the heterostructure. We include not only charge transfer but also the effects of strain by extrapolating the changes in bond length from the flake calculations. The benefit of the MINT representation is that it is a system that effectively models mismatched interfaces that is, nevertheless, well suited to all standard {\em ab initio} methods with periodic boundary conditions. For example, one can carry out the {\em ab initio} total-energy calculations for different magnetically ordered states to probe magnetism. We can also calculate the electronic structure of the MINT representation to obtain effective models for analyses suited to correlated electron methods. 

\emph{Computational methods --- } 
All {\em ab initio} calculations 
were carried out within the total-energy plane wave density-functional pseudopotential approach, using Perdew-Burke-Ernzerhof   generalized gradient approximation functionals\cite{gga} and optimized norm-conserving Vanderbilt  pseudopotentials in the SG15 family\cite{SG15}. Plane wave basis sets with energy cutoffs of 30 hartree were used to expand the electronic wave functions. We used fully periodic boundary conditions and a single unit cell of RuCl$_{3}$ with a $6 \times 4 \times 1$ $k$-point mesh to sample the Brillouin zone. Electronic minimizations were carried out using the analytically continued functional approach starting with a LCAO initial guess within the DFT++ formalism\cite{minimization}, as  implemented  in the open-source code JDFTx\cite{JDFTx} using direct minimization via the conjugate gradients algorithm\cite{conjgrad}. All unit cells were constructed to be inversion symmetric about $z=0$ with a distance of $\approx 60$ bohr between periodic images of the RuCl$_{3}$ surface, using coulomb truncation to prevent image interaction.

\begin{figure}[h]
\begin{centering}
\includegraphics[width=.45\textwidth]{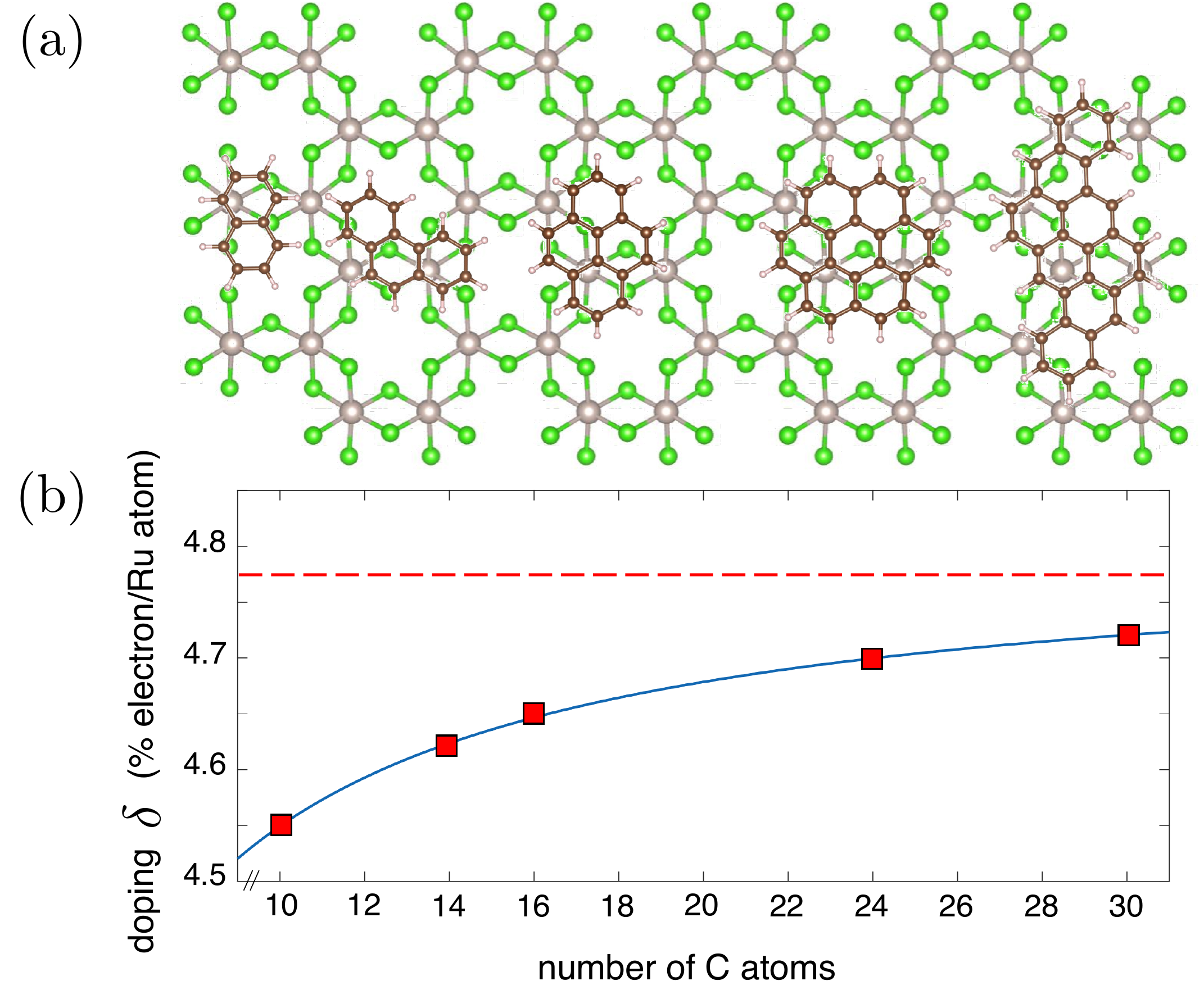} 
\end{centering}
\caption{(a) Visualization of the RuCl$_{3}$-carbon systems considered in our study (top view). (b) Convergence of predicted doping $\delta$ 
of the avatar heterostructures with successively more ``graphenelike" clusters. The red dashed line represents the value of $\delta$ extrapolated from the cluster calculations. Red boxes are the results from calculation and the solid curve is power law fit.} 
\label{fig:dopingc}
\end{figure}

\emph{Application of \emph{MINT} to graphene/$\alpha$-RuCl$_3$ ---} Here, we consider $\alpha$-RuCl$_3$ as the system of interest $S$ and employ hydrogen-terminated graphene clusters $C$ [planar CH$_{10}$H$_{8}$, C$_{14}$H$_{10}$, C$_{16}$H$_{10}$, C$_{24}$H$_{12}$, and C$_{30}$H$_{16}$ as in Fig.~\ref{fig:dopingc}(a)]. To calculate the expected charge transfer in the macroscopic system, we first determine charge transfer for each element of the convergence sequence and then scale to the transfer expected for a full graphene layer $L$ by multiplying by $N(L)/N(C)$, the ratio of the (incommensurate) number of carbon atoms expected for a full graphene layer $N(L)$ and the number in each cluster. Figure~\ref{fig:dopingc}(b) shows that the intrinsic quantity $\delta$ (scaled charge transfer per Ru atom in $S$) converges reliably and rapidly to a value of about 4.77\% $e$/Ru (electrons per ruthenium). 

To test the sensitivity of the calculation with respect to the relative displacements, we have integrated over different relative displacements for a heterogeneous incommensurate bilayer for the first time in a full density-functional theory context. As a matter of expediency, we considered the CH6 cluster and sampled the primitive $d=2$-dimensional surface cell of the $\alpha$-RuCl$_3$ with 12 CH6 configurations: six positions in the plane covering the unit cell and two rotations of the CH6 at each position (see Supplemental Material). We found the averaged doping to be 4.41\%$e$/Ru compared to our original estimate for CH6 of 4.57\%$e$/Ru. This demonstrates that convergence with respect to relative displacement is extremely rapid in this system. Consequently, for all results below, we sample a single relative displacement only.

\begin{figure}[t]
\begin{centering}
\includegraphics[width=.4\textwidth]{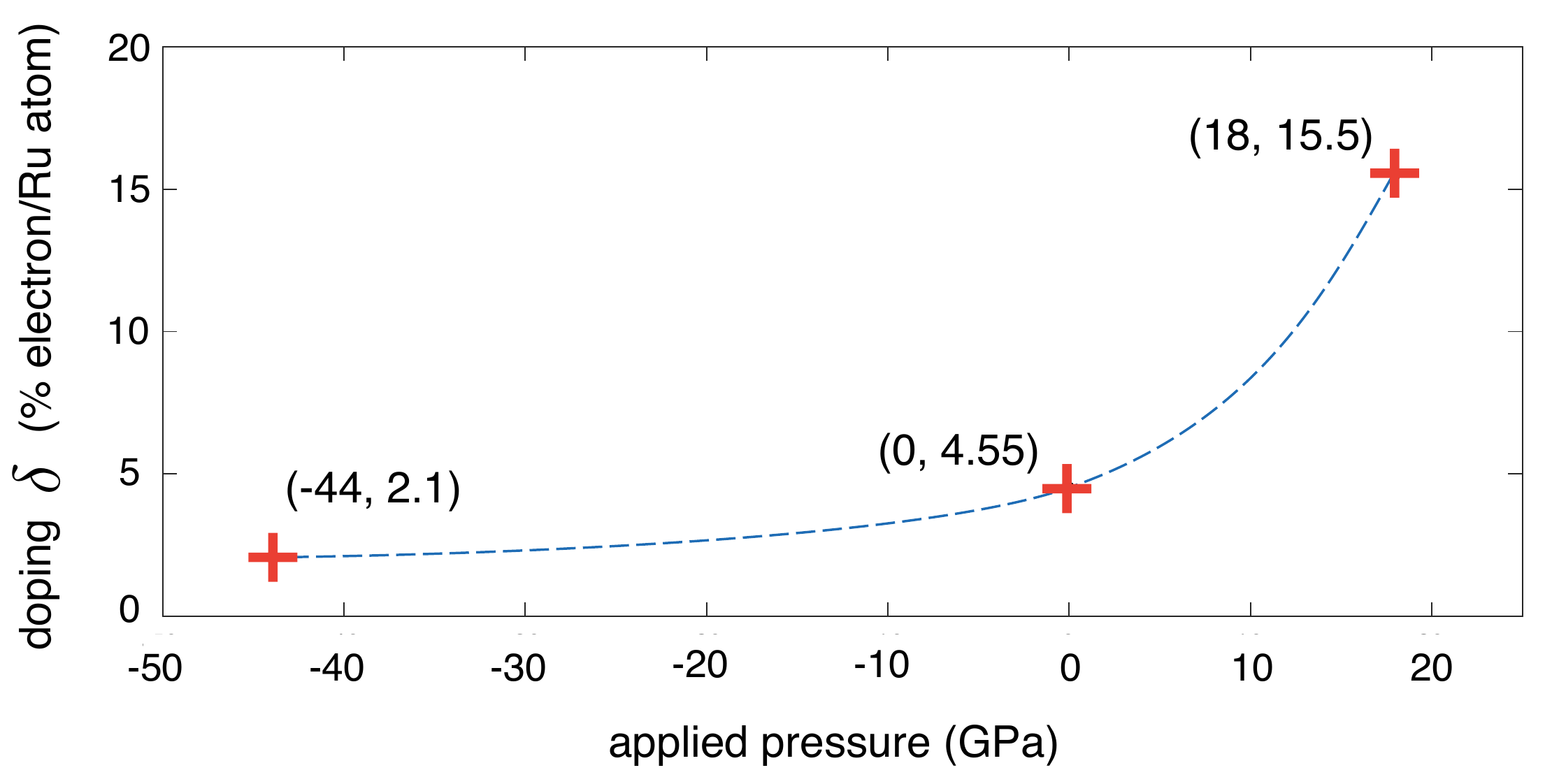} 
\end{centering}
\caption{Doping as a function of vertical  pressure, calculated for the case of a  C$_{10}$H$_{8}$ molecule above monolayer RuCl$_{3}$. The blue curve is a guide for the eye.} 
\label{fig:dopingp}
\end{figure}

We next explore the effect of uniaxial pressure on charge transfer. Holding the $C$ cluster at fixed distance from the $\alpha-$RuCl$_3$ system
 and computing pressure from the resulting force per unit area, we find that $\delta$ increases monotonically with compression and is much more responsive to positive than negative pressure (Figure~\ref{fig:dopingp}). (For these data we used CH$_{10}$H$_{8}$, as it already exhibits good convergence.) We find perpendicular pressure to be an effective tool for controlling doping of GR/RuCl$_3$. 

\textit{Effects on magnetism ---} 
To understand the effect of heterostructuring on magnetism, we consider the charge transfer predicted by MINT added to a single layer of $\alpha$-RuCl$_3$ placed in an effective medium of dielectric constant $1$ that models screening with a debye length 3 \AA , as implemented within JDFTx\cite{dielectric,JDFTx}. This allows us to simulate the doping due to graphene while suppressing unwanted interaction between $\alpha$-RuCl$_3$ planes. We first perform the {\em ab initio} total-energy calculations for the different magnetic ground states. 

\begin{figure}[t]
\begin{centering}
\includegraphics[width=.4\textwidth]{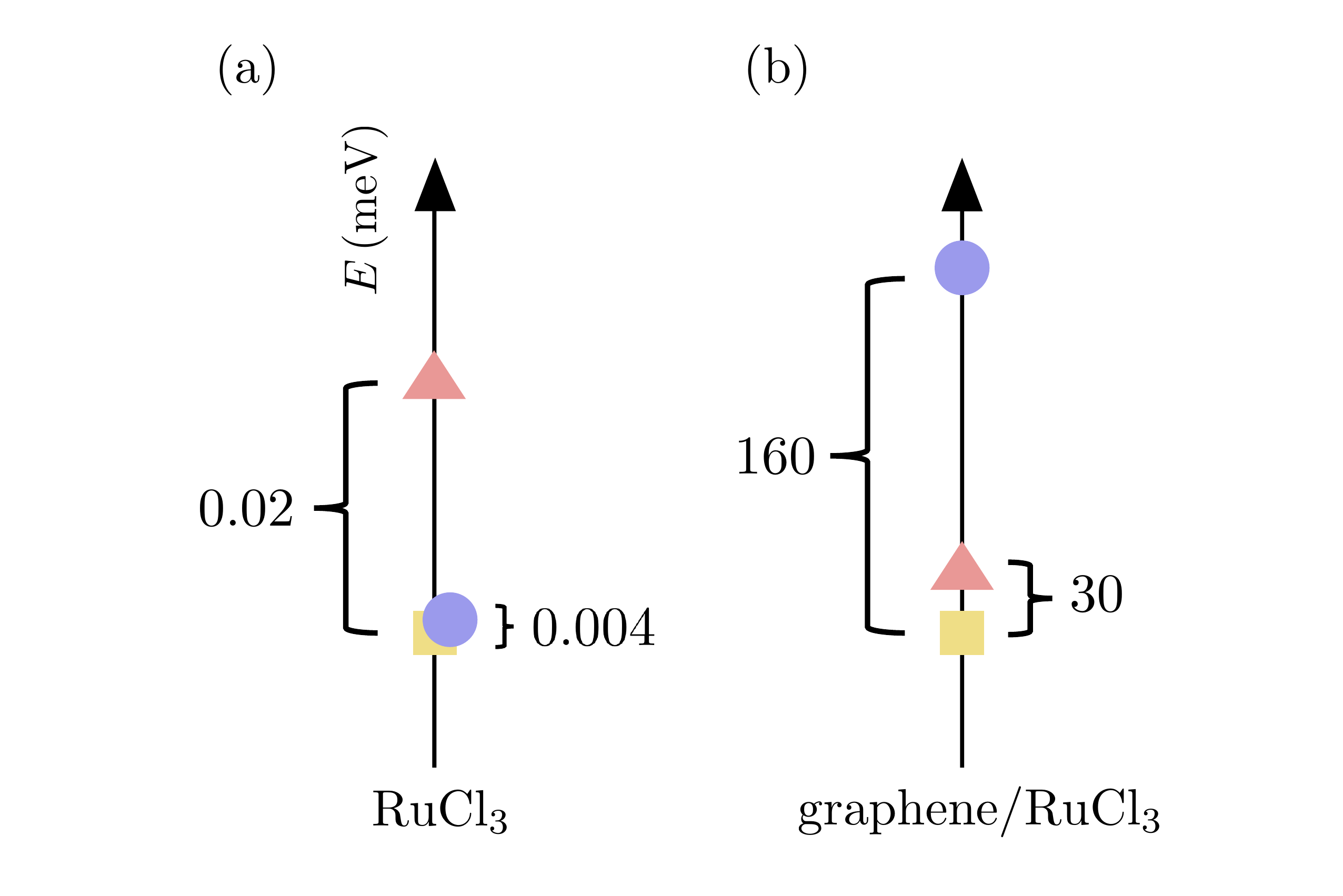} 
\end{centering}
\caption{The energy hierarchy among 
the zigzag- (yellow square), ferromagnetic- (blue circle), and antiferromagnetic-ordered (red triangle) states.}
\label{fig:mag}
\end{figure}

For pure $\alpha$-RuCl$_3$, various \textit{ab initio} studies taking spin-orbit coupling into account have found that the two lowest energy states are ferromagnetic- and zigzag-ordered states that are extremely close in energy. Indeed we find the energy difference between the ferromagnetic state and the zigzag state to be far less than our energy resolution for  pure $\alpha$-RuCl$_3$ (see Fig.~\ref{fig:mag}(a)). On the other hand, the MINT representation for GR/RuCl$_3$ displays a dramatic change in this energy hierarchy. First, the ferromagnetic state experiences a large increase in relative energy. Second, the antiferromagnetic state comes closer to the zigzag state, which remains the lowest energy state (see  Fig.~\ref{fig:mag}(b)).  These results indicate that GR/RuCl$_3$ 
should be closer to the AFM state in the phase diagram of the effective model in Eq.~\eqref{eq:RuCl3-H}.

The full description of GR/$\alpha$-RuCl$_3$
requires understanding of how charge transfer affects the interatomic overlaps that enter the strong coupling expansion of the Kanamori Hamiltonian\cite{Kim-Kee-PD-PhysRevB.91.241110} that results in the magnetic Hamiltonian of Eq.\ref{eq:RuCl3-H}. In addition, one should consider how to describe the doped magnetic system with now more nontrivial Hamiltonian. This second step is beyond the scope of this Letter. Nevertheless we will investigate how the parameters of the magnetic interactions $J$, $K$, and $\Gamma$ are expected to change. 
To accomplish this, we extract the the tight-binding parameters (intra-$t_{2g}$ and nearest-neighbor $t_{2g}-e_{g}$ orbital overlaps) from our \textit{ab initio} calculations on the MINT representation using the maximally localized Wannier orbital method\cite{wannier}. We estimate the on-site Coulomb interaction $U$ following Ref.\cite{PhysRevB.87.165122}. We then use the expressions for the coupling constants in terms of these parameters given in Ref.~\cite{PhysRevLett.112.077204}. The resulting estimates of the NN exchanges give $J/K \simeq -0.3$ and $\Gamma/K \simeq 0.3$ for the doped system, corresponding to the green star on the Luttinger-Tisza phase diagram in Figure \ref{fig:ideal} (b)\cite{Kim-Kee-PD-PhysRevB.91.241110}.
When compared to the previously obtained values of $J/K \simeq -0.7$ and $\Gamma/K \simeq 0.7$ for plain RuCl$_{3}$, this clearly indicates that the charge transfer from graphene to RuCl$_{3}$ has moved the system closer to the Kitaev point.\footnote{This calculation was repeated at an intermediate doping of $\delta=$2.30\% e/Ru to illustrate the trend in magnetism as doping increases. At this value of $\delta$, the parameters extracted via the maximally-localized Wannier-orbital method give $\Gamma/K \simeq 0.4$ and $J/K \simeq -0.5$.}

{\em Summary and outlook --} 
In summary, we have introduced MINT: a new framework for studying lattice-mismatched atomic heterostructures {\em ab initio}. It is a two step process of (1) constructing the MINT representation by combining the cluster and supercell methods to exploit the principle of nearsightedness of electronic matter  and then (2) using this representation to study the electronic structure of mismatched interfaces with full self-consistency in the description of charge transfer across a heterogeneous incommensurate bilayer. We then applied MINT to the GR/$\alpha$-RuCl$_{3}$ system that has recently been realized, finding results that quickly converge to the overall doping of $4.77\%$ electron per Ru atom. A rough estimate based on the work-function difference yields a doping level that is about half our prediction\footnote{This estimate of $2.5 \%$e$/\textnormal{Ru}$ comes from integrating the linear density of states of graphene up to the workfunction difference of $\Delta \phi=1.5$~eV.}.
This rapid convergence provides an internal check of how well the members of the MINT sequence converge to the full GR/$\alpha$-RuCl$_3$ system.
We also predict this doping to increase readily under positive perpendicular pressure.
Finally we predict the doping to bring the GR/$\alpha$-RuCl$_3$ system much closer to the Kitaev point in the phase diagram in terms of effective exchange parameters. 
Interestingly the enhancement in conduction observed in two recent experiments~\cite{Henriksen, Mashhadi} is consistent with our prediction.

The implications of our results are twofold. First, we presented the first framework for studying mismatched interfaces in a systematic yet efficient manner. Although the MINT is a new framework, it is based on a simple principle and it uses established and widely available standard {\em ab initio} methods in each of its steps. Hence MINT is versatile and accessible and we anticipate the application of this approach to produce many more interesting results in mismatched interface systems previously out of reach of {\em ab initio} studies.  Second, using MINT we found the GR/$\alpha$-RuCl$_3$ system to accomplish two sought-after controls: bringing the RuCl$_3$ closer to the Kitaev point and doping. 
To the best of our knowledge, this is the only known case of control that can make the elusive quantum spin liquid physics accessible to RuCl$_3$ without a magnetic field. Our results lay the field wide open to future experiments on GR/$\alpha$-RuCl$_3$ to be compared to MINT predictions. Moreover, it will be interesting to study other heterostructures involving $\alpha$-RuCl$_3$ partnered with different van der Waals systems and continue to explore this uncharted territory.

\begin{acknowledgements}
We thank Leon Balents, Felipe da Jornada, Tim Kaxiras, Sam Lederer, Allan MacDonald, Erik Henriksen, and Ken Burch for useful discussions. 
E-AK was supported by the National Science Foundation (Platform for the Accelerated Realization, Analysis, and Discovery of Interface Materials (PARADIM)) under Cooperative Agreement No. DMR-1539918 and EG was supported by the Cornell Center for Materials Research with funding from the NSF MRSEC program (DMR-1719875).
\end{acknowledgements}

\bibliography{graphenebib}

\clearpage 
\newpage 
\begin{widetext}

\section*{Supplemental Material}

\subsection{Graphene/hBN}
We applied MINT to graphene/hexagonal Boron Nitride (GR/hBN, shown schematically in Figure~\ref{fig:hBNg} (a)) and compared the result with a large, commensurate graphene/hBN supercell in which the graphene lattice constant is initially set to match that of hBN and allowed to relax. The red dashed line in Figure~\ref{fig:hBNg} (b) indicates the extrapolated value of charge transferred from graphene to hBN to be 0.073\% electron/B atom (which is minimal, as expected, since hBN is relatively inert). The value obtained for the commensurate supercell is 0.071\% electron/B atom, thus the two values agree to within 2.77\%, an impressive agreement for such a subtle effect.

\begin{figure}[h]
\begin{centering}
\includegraphics[width=.5\textwidth]{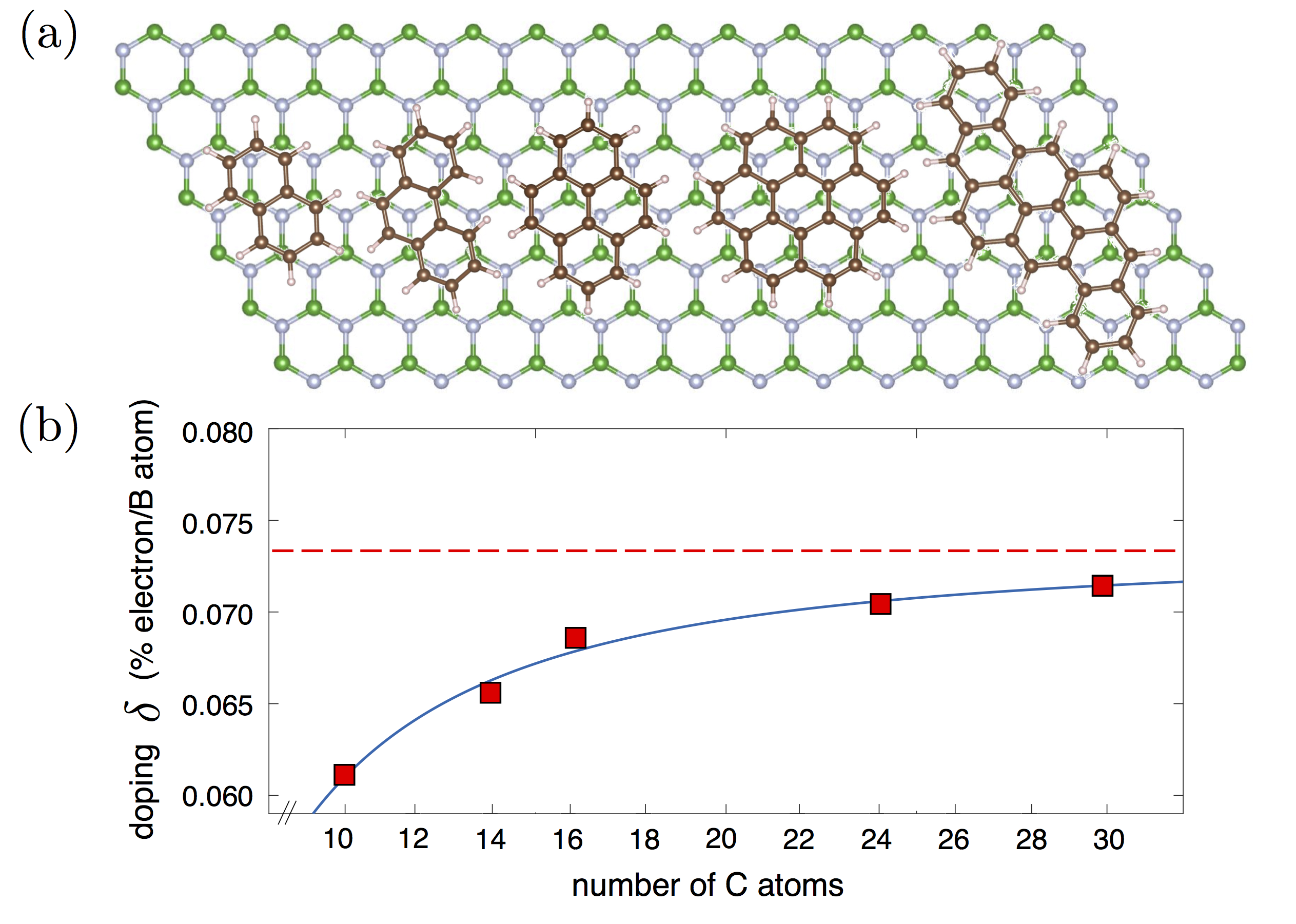} 
\end{centering}
\caption{(a) Visualization of the hBN-carbon systems considered in our study (top view). (b) Convergence of predicted doping $\delta$
of the avatar heterostructures with successively more ``graphene-like" clusters. The red dashed line represents the value of $\delta$ extrapolated from the cluster calculations.} 
\label{fig:hBNg}
\end{figure}

\subsection{Displacement Averaging}
In sampling the primitive $\alpha$-RuCl$_3$ cell, we averaged charge transfer over 12 spatial configurations, shown in Figure~\ref{fig:benzsample}. We considered six CH6 positions in the plane covering the unit cell (three along the long edge of the rectangular unit cell as in Figure~\ref{fig:benzsample} (a) and (c), and three down the middle of the unit cell as in Figure~\ref{fig:benzsample} (b) and (d)) and two rotations of the CH6 at each position. We found the averaged result to be 4.41\%e/Ru compared to our original estimate from CH6 of 4.57\%e/Ru. This close agreement indicates rapid convergence with respect to relative displacement.

\begin{figure}[h]
\begin{centering}
\includegraphics[width=.5\textwidth]{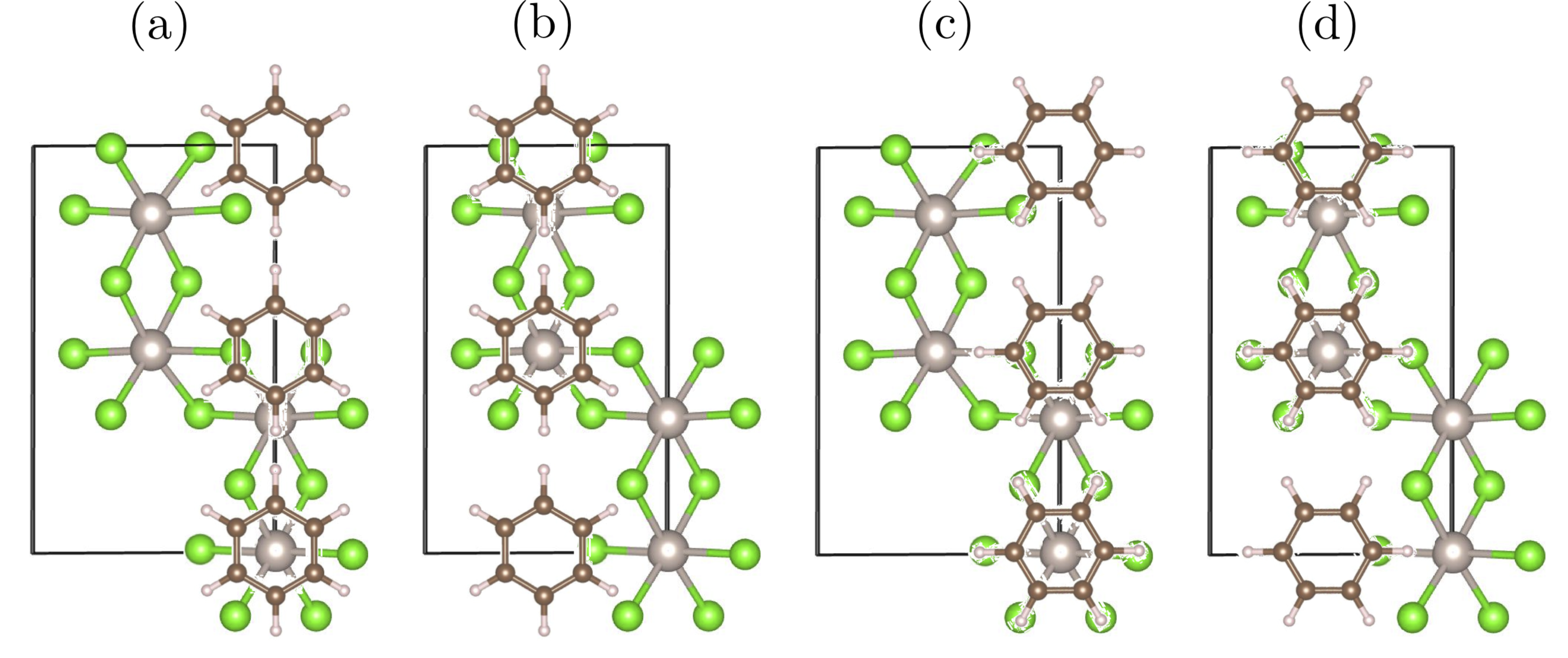} 
\end{centering}
\caption{Spatial configurations considered in our ``displacement-integrated" result for charge transfer between CH6 and $\alpha$-RuCl$_3$. The $\alpha$-RuCl$_3$ unit cell is indicated by the black rectangle, and each CH6 molecule represents a configuration considered in the average.} 
\label{fig:benzsample}
\end{figure}\clearpage

\end{widetext}
\end{document}